\documentclass[printer]{aa}  
\usepackage{psfig}
\usepackage{graphicx}
\usepackage{natbib}

\usepackage{txfonts}

\begin{document}

\title{Dynamical mass of a star cluster in M83: a test  of fibre-fed
  multi-object spectroscopy}

   \author{S. L. Moll
           \inst{1}\fnmsep\thanks{E-mail:s.moll@sheffield.ac.uk}
           \and
	   R. de Grijs
	   \inst{1}\fnmsep\inst{2}
                 \and
	     P. Anders
           \inst{3}
		 \and
             P. A. Crowther
           \inst{1}
	         \and
	     S. S. Larsen
           \inst{3}
		 \and
	     L. J. Smith
	   \inst{4}\fnmsep\inst{5}
 		 \and
	     S. F. Portegies Zwart
           \inst{6}\fnmsep\inst{7}
}
   \institute{Department of Physics and Astronomy, University of
             Sheffield, Sheffield S3 7RH, U.K. 
         \and
             National Astronomical Observatories, Chinese Academy of
             Sciences, Beijing 100012, China 
         \and
             Sterrenkundig Instituut, Universiteit Utrecht, P.O. Box
             80000, 3508 TA Utrecht, The Netherlands. 
         \and
             Space Telescope Science Institute and European Space
             Agency, Baltimore MD 21218, U.S.A  
         \and
             Department of Physics \& Astronomy, University College
             London, London WC1E 6BT, U.K. 
         \and
             Astronomical Institute `Anton Pannekoek' University of
             Amsterdam, 1098 SJ Amsterdam, The Netherlands 
         \and
             Section Computational Science, University of Amsterdam,
             1098 SJ Amsterdam, The Netherlands
}

 \abstract
     {} 
    {We obtained VLT/FLAMES+UVES high-resolution, fibre-fed
  spectroscopy of five young massive clusters (YMCs) in M83
  (NGC~5236). This  forms the  basis of a pilot study  
  testing the  feasibility of  using fibre-fed
  spectroscopy  to measure the
  velocity dispersions of several clusters simultaneously, in order to
  determine their dynamical masses. 
   In principle, this reduces the telescope time required
   to obtain a 
   statistically significant sample of dynamical cluster
   masses. These can be used to assess the long-term survivability of
   YMCs by comparing their dynamical and photometric masses, which are
   necessary to ascertain the potential evolution of YMCs into
   second-generation globular clusters. 
   }
%
   {We adopted two methods for determining the velocity dispersion of the
   star clusters:  cross-correlating the cluster spectrum with the
   template spectra and  minimising a $\chi^2$ value
   between the cluster spectrum and the broadened template spectra. 
   We also considered both red giant and red supergiant template
   stars. 
   Cluster~805 in M83  (following the  notation of Larsen) was
   chosen as a control to test the 
   reliability of the results obtained by this observational method,
   through a comparison with the results obtained from a standard
   echelle VLT/UVES spectrum obtained by Larsen \& Richtler.
   }
%
   {We find no dependence of the velocity dispersions measured for a
   cluster on the choice of red giant versus red supergiant templates,
   nor on the method adopted. However, we do find that the standard
   deviation of the results obtained with only one method may
   underestimate  the true uncertainty.
   We measure a velocity dispersion of
   $\sigma_\mathrm{los}\,=\,10.2\,\pm\,1.1\,\mathrm{km\,s}^{-1}$ for
   cluster~805 from our fibre-fed spectroscopy.
   This is in excellent agreement with the velocity dispersion of
   $\sigma_\mathrm{los}\,=\,10.6\,\pm\,1.4\,\mathrm{km\,s}^{-1}$
   determined from the standard echelle UVES spectrum of cluster~805.
   Our FLAMES+UVES velocity dispersion measurement gives
   $M_\mathrm{vir}\,=\,(6.6\,\pm\,1.7)\,\times\,10^5\,M_\odot$, 
    consistent with previous results. This value of the virial mass is 
    a factor of $\sim 3$ greater  than the cluster's photometric mass,
    indicating a lack    of virial equilibrium.
    However, based on its effective star formation efficiency, the cluster
    is likely to virialise, and may survive for a 
    Hubble time, in the absence of external disruptive forces.
   Unfortunately, our observations of the other M83 star clusters have
   insufficient signal-to-noise  ratios  to determine robust cluster
   velocity dispersions.}
%
   {We find that reliable velocity dispersions can be determined from
   high-resolution, fibre-fed spectroscopy. 
   The advantages of observing several clusters simultaneously
   outweighs the  difficulty of accurate galaxy background
   subtraction,  providing that the targets are chosen
   to  provide sufficient signal-to-noise
   ratios, and are much brighter than the galaxy background.} 

\keywords{galaxies: star clusters -- galaxies: individual: M83 --
  galaxies: spiral} 

 \maketitle


\section{Introduction}

The production of luminous, compact star clusters is
characteristic of  starburst galaxies. However, these clusters have been
observed in several different types of galaxy, including normal
spirals (see e.g. \citealt{larsen99}, and references therein).
These bright clusters are termed Young
Massive Clusters (YMCs), and are usually defined as those star
clusters that are younger than $\sim 1$~Gyr, more massive than
$\sim\,10^5\,\mathrm{M}_\odot$ and  compact, with half-light
radii of a few to a few tens of parsecs. 
The sizes, luminosities and masses of these YMCs are consistent with
the properties expected of young globular clusters. This led to the
suggestion that YMCs represent globular clusters at an early phase of
their evolution \citep{whitmore93,schweizer93}. This would mean that studying YMCs would be
tantamount to observing globular clusters at the epoch of their
birth.  Since globular clusters are amongst the oldest building blocks
in our Galaxy, and are benchmarks for stellar and galactic evolution,
understanding their formation and evolution is
fundamental in developing our understanding of the fields of
large-scale star formation, as well as how galaxies build up and
evolve.   
Determining whether or not YMCs are proto-globular
clusters, therefore, is a vital  step in gaining insights into
these questions.

The simplest criterion that YMCs need to fulfil in order to evolve
into globular clusters is that they must be able to survive for a
large fraction of a Hubble time. Since a cluster requires sufficient
low-mass stars to survive beyond a few Gyr, the  present-day mass
function (PDMF) of a
cluster can be used to assess its survivability. 
It may be possible to get a handle on the PDMF of a cluster by
comparing its dynamical mass to its photometric mass, which is the mass
predicted by evolutionary synthesis modelling based on the observed
 luminosity of the cluster and assuming an initial mass function (IMF).
 If we assume that the dynamical mass of a cluster
represents its true mass, any discrepancies between the two masses can
be attributed to adopting incorrect assumptions in the photometric
mass determination, such as the cluster IMF. 
Even for a very young cluster, which is not in virial equilibrium,
its virial mass can still be used to assess the survivability of the
cluster. This is done by by determining how far out of virial
equilibrium the cluster is, as quantified by its effective
star-formation efficiency (eSFE; \citealt{goodwin06_bast}).
 Thus, dynamical mass
determinations can potentially be used to test the scenario with
respect to the long-term survivability of  YMCs, 
(e.g. \citealt{ho96a,ho96b}; see \citealt{degrijs07_parm} for an
overview and references therein).

Since the velocity dispersion of a cluster needs to be measured in
order to determine the dynamical mass of  a cluster (see
Section~\ref{sec:veldisp}), high-resolution spectroscopy of the cluster is
needed. Large samples of  dynamical cluster masses need to
be obtained to determine whether  YMCs might be
proto-globular clusters, and this can be very  expensive in terms of
telescope time. Observing several clusters at once with a
high-resolution, fibre-fed spectrograph  limits this
cost. 
However, it is unclear whether a sufficiently accurate galaxy
background subtraction can be achieved from only a few fibres
placed at a different position to the star cluster. 
In this paper, we explore the efficiency of this approach, using
  FLAMES+UVES to observe clusters in M83.
 Similar attempts at velocity dispersion measurements have been
  made with FLAMES+GIRAFFE (e.g. \citealt{mieske08}), which uses an
  order of magnitude more fibres than FLAMES+UVES. 
  Since large numbers of fibres can be dedicated to galaxy background
  observations, background  subtraction is more straightforward through
  the use of a master  background spectrum. 
  This, however, is at the expensive of at least a factor of
  two in spectral resolution, and a significantly reduced spectral
  range. 
  This reduced spectral resolution would, however, not permit the
  measurement of the small cluster velocity dispersions  anticipated
  for YMCs ($\sim\,7-15\,\mathrm{km\,s}^{-1}$).
  Furthermore, \citet{mieske08} observed ultra-compact dwarf galaxies
 (UCDs), rather than clusters. These are much  brighter relative to
 the background, and have a much uniform background  than  a spiral
 like M83.

M83, also known as NGC~5236, is a prime example of a nearby spiral
galaxy hosting several YMCs \citep{larsen04_cat}. At a distance of
$4.5\,\pm\,0.3$~Mpc 
\citep{thim03}, with its face-on inclination and multitude of clusters,
M83 is an ideal site to test whether  fibre-fed
spectroscopy can be used to measure the  velocity dispersions of
several clusters simultaneously. A further advantage of using M83 as
the testing ground for this method is that the dynamical masses of two
of its clusters -- clusters~502 and 805 (nomenclature from
\citealt{larsen04_cat}) -- have already been measured 
from standard high-resolution echelle spectroscopy
\citep{larsen04_rich}. Therefore, these clusters can provide a 
control sample to assess the reliability of the results found from
fibre-fed spectroscopy.

M83 is a metal-rich galaxy, with  a central oxygen abundance of 
$12\,+\,\log{\mathrm{O/H}}\,=\,8.94\,\pm\,0.09$ \citep{bresolin05},
nearly twice the solar value of $12\,+\,\log{\mathrm{O/H}}\,=\,8.69$
\citep{asplund}.  The grand-design spiral has a companion, NGC 5253
\citep{rogstad74}, with both galaxies hosting  intense starburst
activity \citep{calzetti99,tremonti01}.
M83 has a nuclear starburst  and ongoing star formation within its
spiral arms \citep{elmegreen98,harris01} and  has a star formation
rate\footnote{This was
   determined   using the equation of \citet{kennicutt98} and adopting
   the average $\mathrm{H}\alpha$ flux of the galaxy as
   $7.1\,\times\,10^{-11}\,\mathrm{erg\,s}^{-1}$ (Kennicutt, priv.
   comm.) and a representative extinction of $E(B-V)\,\sim\,0.5$~mag
  \citep{hadfield05}.  } 
of $4\,\pm\,1\,M_\odot\,\mathrm{yr}^{-1}$.
The galaxy hosts  in excess of 1000 Wolf-Rayet stars
\citep{hadfield05}.

 This paper is structured as follows. In Section \ref{sec:obs}, we present
 our observations of M83 and subsequent data reduction, as well as 
 photometry of the clusters and galaxy background regions.
In Section \ref{sec:veldisp}, we determine the velocity dispersion of
 cluster~805 and compare it with the results obtained from the
 standard UVES  spectrum of
 \citet{larsen04_rich}. We  discuss  the
 possibility of using fibre-fed spectroscopy  to obtain velocity
 dispersions for  other clusters  in  Section  \ref{sec:dis}. 
 Finally, we  summarise our findings in Section \ref{sec:sum}.


\section{Observations and data reduction}
\label{sec:obs}

We obtained fibre-fed spectroscopy of five clusters
and three regions of galaxy background in M83 with the
Fibre Large Array Multi Element Spectrograph (FLAMES) linked with the
Ultraviolet and Visual Echelle Spectrograph (UVES)
on the  Very Large Telescope (VLT) Kueyen Telescope (UT2) in Chile.

\subsection{Imaging of M83}

\begin{figure}
\centerline{\psfig{figure=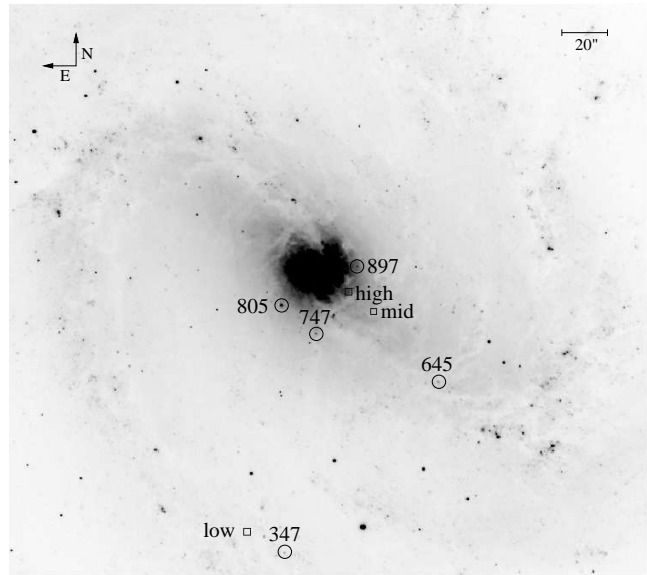,width=8.5cm,angle=0.}}
\caption{\label{fig:m83obs} 
 Archival FORS1 image of M83 with the
 five cluster fibre positions and three galaxy background fibre
 positions indicated. 
}
\end{figure}

\begin{table*}
\begin{center}
\caption
{\label{tab:photom} 
Table of photometry \citep[taken from][]{larsen04_cat}, `background factors' 
and the average background contribution
for the five clusters  observed. 
}
\begin{tabular}{|l| 
ccccc|ccc|c|} \hline
Cluster &\multicolumn{5}{|c|}{Photometry (mag)}     &\multicolumn{3}{|c|}{Background factors} & Background  \\
        &$V$    &$U-B$  &$B-V$  &$V-I$  &$814W$  & Low & Mid & High &contribution ($\%$)\\ \hline
347     &18.27  &-0.62  &-0.06  &0.40   &19.61   & 1.0 & 0.7 & 0.5  &79\\
645     &18.16  &-0.38  &0.23   &0.39   &19.56   & 1.2 & 0.8 & 0.6  &89\\
747     &17.36  &-0.65  &0.25   &0.57   &19.80   & 1.4 & 1.0 & 0.7  &76\\
805     &16.60  &-0.55  &0.19   &0.65   &17.68   & 1.5 & 1.1 & 0.8  &62\\
897     &16.95  &-0.16  &0.51   &0.75   &19.83   & 1.7 & 1.3 & 0.9  &86\\ \hline
\end{tabular}
\end{center}
\end{table*}

We retrieved two archival \emph{I}-band images of M83  taken with the  
Focal  Reducer and low dispersion Spectrograph
 $\#1$ (FORS1) on the VLT. These images, taken as part of ESO
proposal I.D. 60.A-9203(A), were observed  on 9th and 11th March 1999
 UT for 180s each. The \emph{I} filter is
 centred at $7680\textrm{\AA}$ with a $1380\textrm{\AA}$  full-width at
 half-maximum (FWHM), similar to the wavelengths covered by our FLAMES+UVES
spectroscopy.

The images were obtained to aid in selecting the clusters and
regions of galaxy background to observe with the eight fibres of
FLAMES+UVES. 
The clusters were chosen from the catalogue of
\citet{larsen04_cat}. 
First, we selected   cluster~805 as a `control' to assess the accuracy
of our velocity dispersion measurements,  by comparing our
  results with those of \citet{larsen04_rich}.
We then short-listed bright candidate clusters that were denoted in
the catalogue as both `likely/certain cluster'  and having been
successfully  fitted with the 
light-profile fitting program {\sc ishape}  \citep{larsen99}. The
final cluster selection was made on the basis of successfully
positioning the fibres on five clusters and three nearby regions of
galaxy background with the FLAMES fibre-positioning software,
prioritising clusters that were flagged as isolated  with smooth 
  galaxy backgrounds in the catalogue.

The final selection comprised the five clusters 347, 645, 747, 805 and 897
\citep{larsen04_cat} and three regions of galaxy background, as
indicated in Fig.~\ref{fig:m83obs}. The galaxy background
regions were observed so that they could be used to subtract  
the galaxy background from the cluster spectra. Therefore, these had a
range of  intensities, similar  to the galaxy backgrounds
adjacent to the selected clusters.  
The `high' fibre was positioned on a region with a strong galaxy
background, the `mid' fibre was positioned on a mid-intensity
galaxy background region, while the fibre labelled `low' was
positioned on a region with a faint galaxy background.

The FORS1 images were also used  to  assess the relative fluxes of the
 galaxy background regions  adjacent to the  clusters and within
 the galaxy background fibres. 
The images were debiased and flat-fielded with standard calibration
frames. We calibrated the astrometry of the images in  Starlink's 
{\sc  gaia}, using the co-ordinates of the clusters given in the catalogue
of \citet{larsen04_cat}, which were obtained from \emph {Hubble Space
Telescope (HST)} imaging. We carried out our aperture photometry on
both images, also using {\sc  gaia}. We considered the regions of each
galaxy background fibre as well as the regions of galaxy
  background adjacent to each cluster using several small apertures,
with radii of both one and two pixels. Taking a mean of these 
  galaxy background values, we computed the `background factor' for each
cluster and galaxy background fibre pair from the relative 
  galaxy background fluxes measured. This background
factor is the factor by which the galaxy background flux in the
 galaxy background fibre needs to be multiplied in order to 
(approximately) correspond to the galaxy background adjacent to the
considered cluster. These background factors are contained in
Table~\ref{tab:photom}, along with  photometry of the clusters taken
from \citet{larsen04_cat}.

\subsection{Spectroscopy of M83}
\label{sec:m83spec}

We observed the five clusters and three galaxy background regions of
M83 indicated 
in Fig.~\ref{fig:m83obs} over several nights in service mode in 2006,
using  FLAMES+UVES.
The  average seeing was $\sim 1$~arcsec. 
When linked with UVES, FLAMES is a multi-object, high-resolution
spectrograph ($R=47~000$) with eight fibres of 1~arcsec diameter
aperture. 
In  order to position the fibres, each of the eight objects that can
be observed at one time must have at least a 10.5~arcsec separation.
There are two plates on the fibre positioner, so 
that the fibres can be positioned on one plate while observations are 
taking place on the other, thus reducing acquisition time. Each plate
has a different spectral response 
and a slightly different wavelength calibration.   
Although UVES is a cross-dispersed echelle
spectrograph, with both a blue and red arm that can obtain data
simultaneously, only the standard red UVES set-ups (centred on 5200,
5800 or $8600\textrm{\AA}$) can be used with FLAMES. This red arm
comprises a mosaic of an EEV CCD for the shorter wavelengths and a
MIT-LL CCD for the longer wavelengths.
We observed for a total exposure time of $22~048$~s, using cross
disperser $\#4$, centred at $8600\textrm{\AA}$  with 
$312\,l\,\mathrm{mm}^{-1}$, thus obtaining data from
{$\sim 6670-10425\,\textrm{\AA}$}. 
The region {$\sim 8545-8645 \,\textrm{\AA}$} was not
observed because it lay within the gap between the two CCDs. 
 ThAr arcs
were taken for wavelength calibration. The resolution of the data, as
measured from the ThAr arcs, is $\sim 9\,\mathrm{km\,s}^{-1}$. 
The unbinned
data  have a pixel scale of 0.182~arcsec\,pixel$^{-1}$.

Since the spectra of M83 were faint,  we decided that it was more
appropriate to reduce the data by hand, rather than to use the
pipeline-reduced data. The data were reduced using the software
package {\sc iraf}, considering the data from each night, and, where
relevant, each fibre-positioning plate  separately.
First, a median bias frame for each night was produced and
subtracted from the data.
We then divided each data frame by a master slit  flat-field for
each night.
This was produced by taking a median of the nine slit flat-fields that
had been taken in three different slit positions each night, rejecting the
six faintest frames, and normalising.
By rejecting the six faintest frames, the pixels that lay in any of the 
inter-order gaps in the slit  flat-fields were rejected, so that most
of the chip was illuminated in the master slit flat-field. All data
 lay on the illuminated part of this master slit flat-field.
The next step was to divide by `fibre flat-fields', which are
flat-fields that are obtained through the fibres, and can be used to
correct for the relative response of each fibre.
We divided by the normalised 2-D all-fibre flat-field. This compared
favourably with the results achieved by 
dividing the extracted data 
by the fit to the response of extracted odd- and even-fibre flat-fields.
All eight fibres of the data were then optimally extracted \citep{horne86},
each with an  aperture of 1.456 arcsec (8 pixels) and wavelength
calibrated 
with the extracted ThAr arcs. We determined the heliocentric radial
velocity for each night's data using the Starlink program {\sc rv}
\citep{rvref} and
corrected with these values. The data for each fibre were then co-added,
and the orders merged. 
The 1-D data were then galaxy-background subtracted. This was
done by subtracting an 
average galaxy background spectrum, produced from the  two 
  galaxy background fibres with sky factors 
closest to unity (see  Table~\ref{tab:photom}), after multiplication
by these sky factors. 
The relative contribution of the galaxy
background to the cluster\,+\,galaxy background spectrum
is also included in Table \ref{tab:photom}.

We measured a recessional velocity of
$v_r\,=\,500\,\pm\,4\,\mathrm{km\,s}^{-1}$,  after heliocentric radial
velocity correction, from the observed  central
wavelengths of the two  Ca~{\sc ii} triplet lines visible in our data 
($8498.02\,\textrm{\AA}, 8662.13\,\textrm{\AA}$) for cluster~805.
 These central wavelengths were found by
fitting the lines with the  {\sc elf} (emission-line fitting)
  routine   in Starlink's {\sc dipso} package \citep{dipsoref},
allowing line width, centre and intensity to vary freely.
The recessional velocity implied from the  better signal-to-noise ratio
$8498\,\textrm{\AA}$ line was more heavily weighted, according to the
uncertainty in the fitted central wavelength estimated by {\sc elf}. 
This value is in good agreement with other optical measurements:
  $503\,\pm\,11\,\mathrm{km\,s}^{-1}$ \citep{devauc91},
  $497\,\pm\,18\,\mathrm{km\,s}^{-1}$,
  $445\,\pm\,20\,\mathrm{km\,s}^{-1}$ \citep{fouque92} and
  $491\,\pm\,30\,\mathrm{km\,s}^{-1}$ \citep{humason56}. 
  Since no strong lines, including the  Ca~{\sc ii} triplet, could
  be identified in the spectra of any cluster except cluster~805, it
  was not possible to determine individual cluster recessional
  velocities. Therefore, even though small differences in radial
  velocity may exist between the clusters, all cluster spectra were
  velocity corrected  by $v_r\,=\,500\,\mathrm{km\,s}^{-1}$.
The spectra were then normalised, by fitting a polynomial to the
 continuum across the whole spectrum and dividing by this fit.

\subsection{Spectroscopy of standard stars}
\label{sec:data}

With the setup described above (Section \ref{sec:m83spec}), we also
observed a telluric star, 
CD-32$^\circ$9927, and three  Galactic red giants -- HD~159821 (K1~III),
HD~123833 
(K2~III) and HD~159881 (K5~III).  We chose to observe giants rather
than supergiants, because these four calibration stars could be observed in
only two fields,  reducing the
overheads of the observations. For each of these standard stars, one
sky
fibre was placed as close to the object as possible.

We chose to use the pipeline-reduced data for these bright standard
stars, since these included a better extraction of  the 
$8662\,\textrm{\AA}$  Ca~{\sc ii} triplet lines than the
manually-reduced data.  This line lies very close to
the edge of the MIT-LL CCD for the template stars. Due to their
higher recessional velocity, this is not a
problem  for the M83 clusters.
We sky  subtracted the spectra of these bright
standards by simply subtracting the relevant sky
  fibre spectrum. 
The data were heliocentric radial velocity-corrected using
the values computed by {\sc rv}. 
Each red giant was then corrected for its recessional
velocity using the observed central wavelengths of
the two  visible Ca~{\sc ii} triplet lines, as measured with {\sc elf}.
 The spectra were then normalised.

\begin{figure*}
\centerline{\psfig{figure=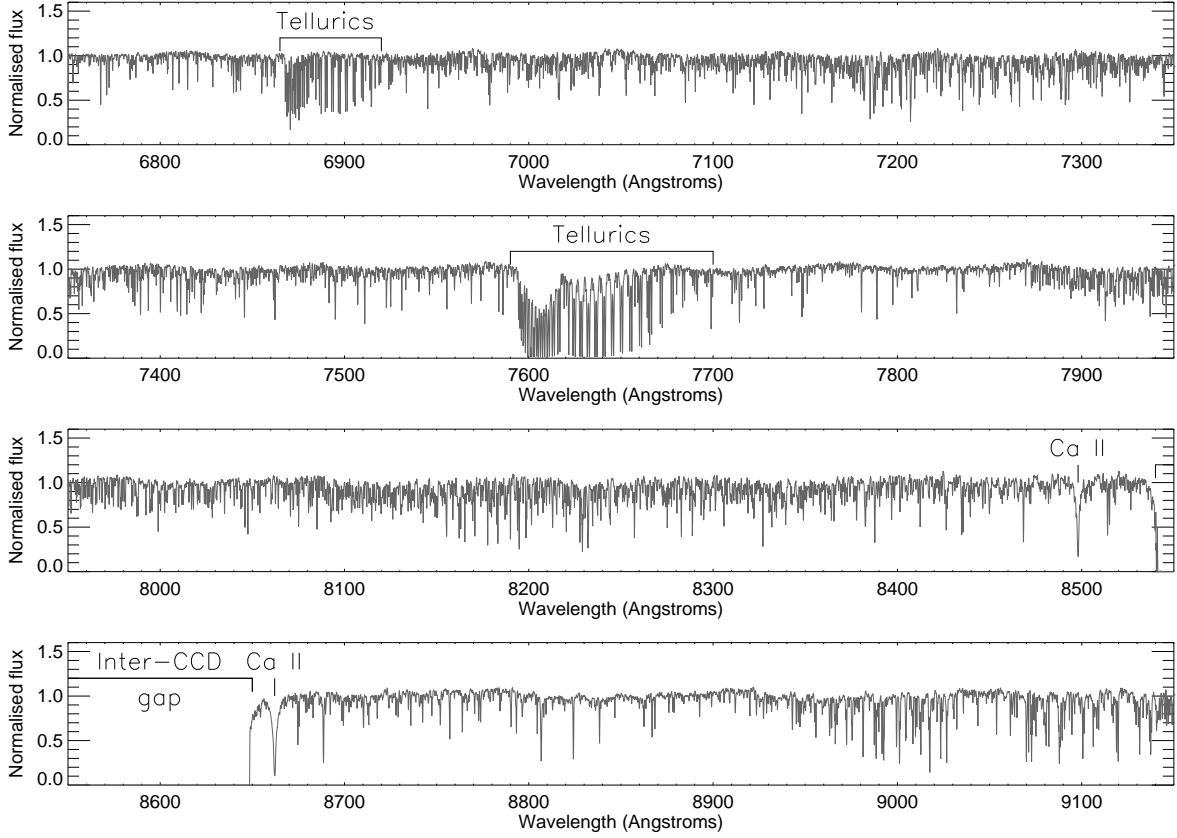,width=16.cm,angle=0.}}
\caption{\label{fig:temp} 
Pipeline-reduced spectrum of the K1 giant HD~159821, normalised and
corrected to rest wavelength, over the wavelength range
$6750-9150\,\textrm{\AA}$. The two lines  of the Ca~{\sc ii} triplet that
are contained in our data ($8498.02\,\textrm{\AA},
8662.13\,\textrm{\AA}$) are marked. The other line of the Ca~{\sc ii}
triplet ($8542.09\,\textrm{\AA}$) lies on the gap between the two CCDs,
which is also indicated.  The
major telluric features in this spectral range are also marked.
}
\end{figure*}

\begin{figure*}
\centerline{\psfig{figure=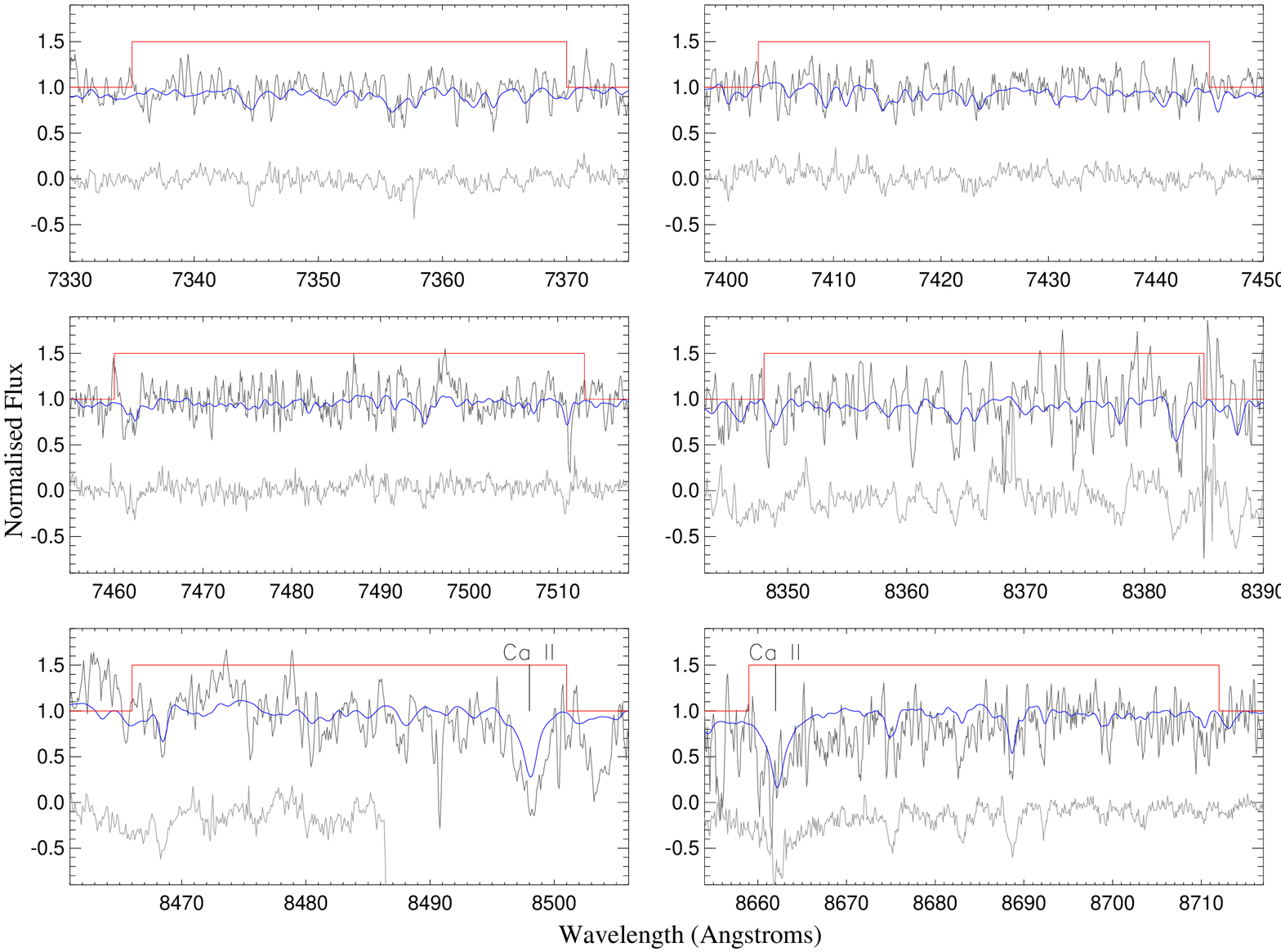,width=17.cm,angle=90.}}
\caption{\label{fig:veldisp} 
Best fit of broadening for the K2~III template HD~123833 (blue line)
to our normalised FLAMES+UVES cluster~805 spectrum (black line) 
  for the `Both CaT' case of
$10.9~\mathrm{km\,s}^{-1}$. 
The spectral regions included in the fit are shown where the red line
is at 1.5. 
Also included in this figure is the standard echelle UVES spectrum of
cluster~805 from \citet{larsen04_rich} (grey line), offset
at -1. 
See the online journal for a colour image. 
}
\end{figure*}

An example spectrum of a velocity-corrected, normalised red giant
template star, HD~159821 (K1~III) is shown in Fig.~\ref{fig:temp}, for
a reduced wavelength coverage.
The principal telluric features in
this spectral range and the two visible Ca~{\sc ii} triplet lines are
marked.
 An example cluster spectrum over suitable spectral regions is shown
 in Fig.~\ref{fig:veldisp}, fitted with a red giant template star.


\section{Virial mass}
\label{sec:veldisp}

The virial mass of a cluster, $M_{\mathrm{vir}}$ is given by the
equation:
$$
\label{eqn:M}
 M_{\mathrm{vir}} \approx 
     \frac {\eta \sigma_\mathrm{los}^2 r_{\mathrm{hl}}}{G}~,
$$
where $\sigma_\mathrm{los}$ is the line-of-sight velocity dispersion
of the cluster 
and  $r_{\mathrm{hl}}$ is the projected half-light radius of the cluster. This
equation follows from the virial theorem, with $\eta=9.75$ for a wide
range of light profiles, assuming that the cluster is virialised,
gravitationally bound and spherically symmetrical, that the velocity
dispersion of the cluster is isotropic and that all of the stars
contained in the cluster are single stars of equal mass
\citep{spitzer87}. 
However, recent studies have shown that $\eta$ is not a constant, and
 can vary with time, depending 
 on, for example, the degree of mass segregation  and the binary fraction of
 the cluster (\citealt{boily05,fleck06,kouwenhoven08}). 
 So that our results are comparable with other dynamical mass
  determination of YMCs, we adopt $\eta=9.75$, as has been done by
  other authors.

\subsection{Velocity dispersion}

Clearly, it is necessary to measure the line-of-sight velocity
dispersion of the cluster in order to determine the cluster's virial
mass. This is achieved by comparing the widths of the
Doppler-broadened lines of the cluster with a template star.
The cluster spectrum is compared to either red giant or red supergiant
template star spectra, because the lines of these stars are
broadened by only a few  $\mathrm{km\,s}^{-1}$ by macro-turbulence in
their atmospheres \citep{gray86}. It is not appropriate to use earlier type
supergiants and main sequence stars, as effects such as rotational
broadening, macro-turbulence and micro-turbulence broaden the lines
of these stars by amounts  substantially greater than the
anticipated cluster 
velocity dispersions.  
Therefore, only spectral regions redwards of
$\sim 5000 \textrm{\AA}$, which are dominated by light arising from
cool supergiants of spectral type F-M, should be considered (see
e.g. \citealt{ho96a}). 
 Red supergiant templates are preferable, as young clusters do not
contain red giants. However, red giants can be used, although these have 
lower values of macro-turbulent broadening than more 
luminous giants. This could mean that comparing the spectrum of a cluster
that is dominated by red supergiants with a red giant template spectrum
may systematically overestimate the cluster velocity dispersion.
However, \citet{larsen04_brod} found no systematic differences, within
the uncertainties, in
the velocity dispersions determined for a cluster when
cross-correlating with templates of luminosity classes I, II and
III.

Here, we adopted two methods to determine the line-of-sight velocity
dispersion. The first involved minimising a  $\chi^2$ value between
the cluster spectrum and broadened template spectra.
The second involved cross-correlating the cluster spectrum with each
template spectrum. These two methods are discussed in more detail in
Sections \ref{sec:chisq} and \ref{sec:xcor}.

To make a direct comparison between the results that can be obtained
with fibre-fed spectroscopy and those from a standard echelle
spectrum, we also determined velocity 
dispersions using both  methods with the non-fibre-fed UVES
spectrum of cluster~805 from  \citet{larsen04_rich}.
We degraded the UVES spectrum by re-binning the data to the same
linear dispersion scale as our own. 
To consider the effect of red supergiant over red giant templates we
also included two red supergiant template stars  from the
UVES Paranal Observatory Project (POP; \citealt{POPref}): HD~206778~(K2~Ib) and HD~12642~(K5~Iab). These were also
re-binned to the same lower linear dispersion 
scale as our own data.

\subsubsection{ $\chi^2$ minimisation}
\label{sec:chisq}

We measured the line-of-sight velocity dispersion of
cluster~805 by minimising a  $\chi^2$ value between the
cluster spectrum and each broadened template
spectrum in turn. All spectra were normalised and then the
  continuum was subtracted.  
The broadening was achieved by convolving a 
template spectrum with a Gaussian of $\sigma$ equal to the desired
velocity broadening. To find the best fit, the $\chi^2$-minimisation
program considered a range of broadening values, 
over selected spectral regions. The criteria for selection were
  that the spectral regions showed visible similarity in absorption
  lines between the cluster and template spectra and that they did 
  not  contain any telluric lines, which could provide an artificial
  match between the spectra.
These spectral regions  were: 
$7335-7370\textrm{\AA}$, $7403-7445\textrm{\AA}$,
$7460-7513\textrm{\AA}$, $8348-8385\textrm{\AA}$,
$8465-8502\textrm{\AA}$ and $8659-8712\textrm{\AA}$.
Rather than considering these small regions of spectra individually,
we ran the model over the whole spectrum, but only considered the
chosen spectral regions when computing the  $\chi^2$ value. 
Non-linear steps of broadening were considered in the
  $\chi^2$-minimisation program, with closer steps as  the
  minimum was approached.
 An example fit of a broadened template star
spectrum to the cluster~805 spectrum over these spectral regions is
shown in Fig.~\ref{fig:veldisp}.

 Since the Ca~{\sc ii} triplet lines may be saturated in the template
spectra, including them in the fit may  overestimate the cluster
  velocity  dispersion (e.g. \citealt{walcher05,martini04}).
However, \citet{mengel02} found no disparity between the velocity 
dispersion results computed by $\chi^2$  minimisation for the strongest
component of the Ca~{\sc ii} triplet and other individual absorption
features for clusters in NGC~4038/4039. 
We decided to consider the potential impact of this effect by
excluding these lines from the fit. Therefore, we considered three
cases, with each case including slightly different spectral regions in
the fit.  
The first case used the spectral regions described above, including
both visible  Ca~{\sc ii} triplet lines. Thus, it was called `Both
  CaT'.  
We also considered the case where the $8498\textrm{\AA}$
Ca~{\sc ii} line was masked out of the fit, still including the
$8662\textrm{\AA}$ Ca~{\sc ii} line, since the $8498\textrm{\AA}$ was
not in the UVES spectrum of \citet{larsen04_rich}. 
This case was called `$8662\textrm{\AA}$'.
In the final case, both Ca~{\sc ii} lines were masked (called `No
CaT'). 
The results for all of these cases are given in
Table~\ref{tab:veldisp}.  

\newpage

\begin{table*}
\caption{
\label{tab:veldisp}
Velocity dispersions in $\mathrm{km\,s}^{-1}$ determined by comparing
our FLAMES+UVES cluster~805 spectrum 
and the standard UVES cluster~805 spectrum of \protect
\citet{larsen04_rich} 
with five red giant (RG) and red supergiant
(RSG) template spectra.
}
\begin{center}
\begin{tabular}{|l||cc|ccc||c|c|} \hline
          &\multicolumn{5}{|c||}{$\chi^2$ minimisation}
                     &\multicolumn{2}{|c|}{Cross-correlation} \\ \cline{2-8}
Template          &\multicolumn{2}{c|}{UVES}
                            & \multicolumn{3}{c||}{FLAMES+UVES} 
                                                      &UVES 
                                                              &FLAMES+UVES \\
          &`No CaT'$^\star$&`$8662\textrm{\AA}$'$^\star$   
                            &`No CaT'$^\star$ &`$8662\textrm{\AA}$$^\star$'&`Both CaT'$^\star$ 
                                                   &`No CaT'$^\star$&`No CaT'$^\star$\\ \hline
K1~III   & 10.3 & 10.4     & 9.5  & 9.8  &  9.9    & 9.8  & 9.1 \\ 
K2~III   & 10.8 & 10.9     & 10.2 & 10.6 & 10.9    & 9.8  & 10.0 \\
K5~III   & 11.8 & 11.9     & 10.8 & 11.1 & 10.9    & 8.4  & 8.4 \\ 
K2~Ib    & 13.2 & 13.2     & 12.5 & 12.7 & 12.6    & 9.8  & 10.4\\
K5~Iab   & 11.7 & 11.7     & 10.8 & 11.1 & 11.0    & 10.0 & 10.2\\ \hline
Mean of RGs$^\dagger$ 
 &10.9\hspace{0.2cm}(0.8)  &11.0 \hspace{0.2cm}(0.8)    
   &10.1\hspace{0.2cm}(0.7) &10.5 \hspace{0.2cm}(0.7)  &10.6 \hspace{0.2cm}(0.6)
     & 9.3 \hspace{0.2cm}(0.8)  &9.2 \hspace{0.2cm}(0.8) \\
Mean of RSGs$^\dagger$ 
 &12.4 \hspace{0.2cm}(1.1)  &12.5 \hspace{0.2cm}(1.0)    
  &11.7\hspace{0.2cm}(1.2) &11.9 \hspace{0.2cm}(1.1)  &11.8 \hspace{0.2cm}(1.1)
   & 9.9 \hspace{0.2cm}(0.1)  &10.3 \hspace{0.2cm}(0.1) \\
Mean of all$^\dagger$ 
 & 11.5\hspace{0.2cm}(1.1)  &11.6 \hspace{0.2cm}(1.1)    
  &10.8\hspace{0.2cm}(1.1) &11.1 \hspace{0.2cm}(1.1)  &11.1 \hspace{0.2cm}(0.9)
\    & 9.6 \hspace{0.2cm}(0.7)  &9.6 \hspace{0.2cm}(0.8) \\ \hline
\multicolumn{8}{p{18cm}}{$^\star$This indicates the spectral region
  considered. `Both CaT' includes all
selected spectral regions (see text),  `$8662\textrm{\AA}$' excludes
  the $8498\textrm{\AA}$ Ca~{\sc ii} line and `No CaT' excludes  both
  visible Ca~{\sc ii} lines. }  \\
\multicolumn{8}{p{18cm}}{$^\dagger$The standard deviation on each
  mean is presented in parentheses} \\
\end{tabular}
\end{center}
\end{table*}

The results obtained from this method may be sensitive to low signal-to-noise ratios. 
Taking $\Delta \chi^2 = 1$ as the $1\sigma$ error \citep{press92}, we
estimate that the 
uncertainties on the FLAMES+UVES velocity dispersions are
$\sim 2\,\mathrm{km\,s}^{-1}$, while the uncertainties on  the UVES
velocity dispersions are $\sim 1\,\mathrm{km\,s}^{-1}$.

\subsubsection{Cross-correlation method}
\label{sec:xcor}

In this second method, we determine the cluster velocity dispersion
using the cross-correlation technique of   \citet{tonry79}. All
spectra were normalised and then the continuum was subtracted.
We cross-correlated 
the cluster spectrum with each template spectrum in turn, over the
spectral regions described in Section \ref{sec:chisq} as `No CaT',
excluding the Ca~{\sc ii} triplet lines. 
 We  did not consider the other two cases, since  broadly consistent
results for all three cases were obtained for the $\chi^2$
minimisation.

The FWHM of the resulting cross correlation function (CCF) relates to
the velocity dispersion of the cluster. This relationship was
calibrated by considering the template star used. The template star was
broadened and cross-correlated with the original, unbroadened
template. By considering several values of broadening, and measuring
the FWHM of each CCF, the almost linear relationship between the FWHM
of the CCF and broadening was empirically  calibrated to an
absolute scale, from which the velocity dispersion of the cluster was
read. This was  repeated for each template, because every star has a
different calibration. 

This method is less sensitive to spectral-type matching and low
  signal-to-noise than the $\chi^2$ 
  minimisation. However, it suffers from complications associated with the 
  subjectivity of fitting weak, non-Gaussian CCFs. Based on a $5\,\%$
  error in fitting the FWHMs of the CCFs, each measurement has an
  uncertainty of $\sim 1\,\mathrm{km\,s}^{-1}$. Note that it is
  not the intrinsic FWHM that is important, but the relative FWHMs.

\subsection{Comparison of results and techniques}
\label{sec:comparison}


Firstly, it is apparent from Table~\ref{tab:veldisp} that 
there is no systematic difference
in the velocity dispersion measured when comparing a cluster spectrum
with red giant or red supergiant templates. This is true for both the
$\chi^2$ minimisation technique and the cross-correlation. For
this reason, we will consider the mean of all five templates when
discussing results below.

Secondly, we do not
observe a systematic increase in the velocity dispersion measured when
the Ca~{\sc ii} lines were included in the $\chi^2$ minimisation. 
 This  may  be due to the large wavelength range considered in the  
  fit, thus reducing the weight applied to the strong Ca~{\sc ii}
  triplet lines. This  would also
  explain the poor fitting of these lines.
Due to this similarity in results, we consider the results obtained
for all five template 
stars with the `No CaT'  case of spectral regions for both
techniques. The mean and standard deviation  of these results give  
$\sigma_\mathrm{los}\,=\,10.2\,\pm\,1.1\,\mathrm{km\,s}^{-1}$.

We find that, within the uncertainties the velocity dispersions found
from the FLAMES+UVES data are consistent with those measured from the
standard UVES data, for both methods and all spectral regions
considered. If we also take a mean of the ten measurements obtained
for the two techniques with the `No CaT' case of spectral regions for
the standard UVES data, we find
$\sigma_\mathrm{los}\,=\,10.6\,\pm\,1.4\,\mathrm{km\,s}^{-1}$, again
with the uncertainty representing the standard deviation of the
results. This is in excellent agreement with our FLAMES+UVES
measurement of   
$\sigma_\mathrm{los}\,=\,10.2\,\pm\,1.1\,\mathrm{km\,s}^{-1}$.
 Therefore, we conclude that our results agree with those
obtained from standard echelle spectroscopy, and that
fibre-fed spectroscopy  \emph{is} practicable for the
purpose of determining cluster velocity dispersions.


Our FLAMES+UVES velocity dispersion result is  also in 
agreement with the result measured by 
\citet{larsen04_rich} of
$\sigma_\mathrm{los}\,=\,8.1\,\pm\,0.2\,\mathrm{km\,s}^{-1}$, at the
$2\sigma$ level. 
This was determined  by    cross-correlating their UVES spectrum with
red supergiant  templates.  They considered several spectral regions
separately:   
$4520-4850\textrm{\AA}$, $7310-7570\textrm{\AA}$,
$7710-8070\textrm{\AA}$ and $8680-8880\textrm{\AA}$, as well as
considering two cases including multiple regions: 
the region $6770-6850\textrm{\AA}$ with $6940-7110\textrm{\AA}$ and
the region $6770-6850\textrm{\AA}$ with both $6940-7110\textrm{\AA}$
and  $7310-7570\textrm{\AA}$. 

We only obtained results  for cluster~805.  
  The other clusters, which are fainter and have higher relative
  background contributions, have lower signal-to-noise ratios.
We measure a continuum signal-to-noise ratio per resolution element 
 ($\approx\,0.1\,\textrm{\AA}$) for cluster~805 of $\sim 7$ at 
$7000\,\textrm{\AA}$ and  $\sim 2$ at $9000\,\textrm{\AA}$ compared
with a signal-to-noise ratio per resolution element for cluster~897,
the next brightest cluster in the $I$-band,  of $\sim 6$ at
$7000\,\textrm{\AA}$ and  $\le\,1$ at $9000\,\textrm{\AA}$. 
  For these other clusters, the noise is such  that no spectral
  regions could be found 
  for which  the  absorption lines  present in the template star could
  be identified in the cluster spectrum. Computing the $\chi^2$
  minimisation over the regions adopted for cluster~805 led to a
  broadening that was an order of magnitude too large.
  Cross-correlation over these regions did not produce a peak in
  the CCF. 
  In the case of cross-correlation, the lack of results cannot be due
  to differences in radial velocities between these
  faint clusters. Indeed, had a CCF peak been produced, the radial 
  velocities of these clusters could have been measured from its position.

The rapid decline in the signal-to-noise ratio in the spectra of all
  clusters, including cluster~805,  redwards of the 
  Ca~{\sc ii} meant that we did not include these 
regions in the velocity dispersion determinations.
  If the galaxy background was not subtracted from the spectrum of
  cluster~805, and the sky emission lines were masked out of the fit, 
the velocity dispersion of the cluster could not be
  measured. This is probably due to the fact that the lines were
  significantly diluted by the background. Therefore, this could also
  not be used to measure the velocity dispersion for any of the other
  four clusters.

 Clearly, in order to measure reliable star cluster velocity
dispersions, good signal-to-noise ratios in the cluster spectra are
paramount. 
 We obtained consistent results from cluster~805 with a
 signal-to-noise ratio of  
$\gtrsim\,3$  in the spectral regions considered, but no results from 
the data of the other clusters, with lower  signal-to-noise ratios.
Therefore, a signal-to-noise ratio of $\sim 3$ seems likely to
represent a lower limit to the signal-to-noise ratio required to
obtain reliable results.
To investigate this, we produced a simple model cluster by adding the
library K2~Ib and K5~Iab spectra, re-normalising the spectrum and
broadening it by $10\,\mathrm{km\,s}^{-1}$. A $\chi^2$ minimisation
was carried out between this model spectrum and all five template
stars for the `Both CaT' case of spectral region, 
in an attempt to retrieve this value of the velocity dispersion. 
The signal-to-noise ratio of the model cluster spectrum
was degraded by artificially adding Gaussian noise and the $\chi^2$
minimisation was recomputed for the new noisier model spectrum. This
was repeated for a range of  model cluster spectrum signal-to-noise
ratios.

\begin{figure}
\centerline{\psfig{figure=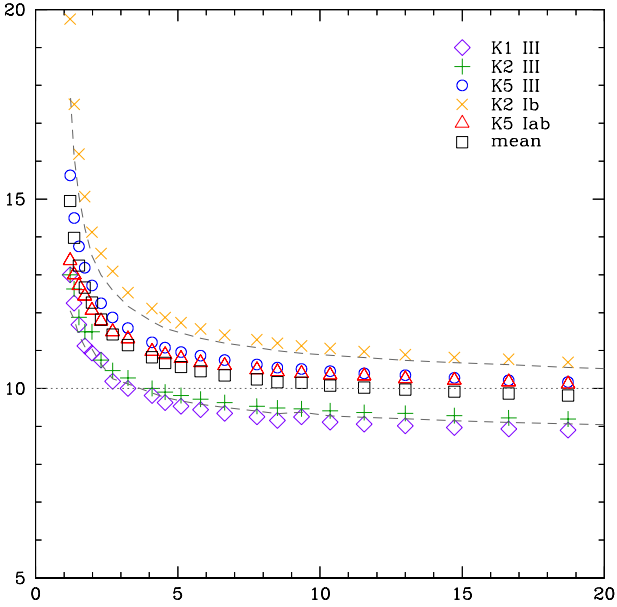,width=0.49\textwidth,angle=0.}}
\caption{\label{fig:snr}
Plot of velocity dispersion measured from a $\chi^2$ minimisation
between the five template stars and a simple model cluster spectrum
with $\sigma_\mathrm{los}\,=\,10\,\mathrm{km\,s}^{-1}$ in the `No CaT'
case against the signal-to-noise ratio  (SNR) of the model cluster.
The mean velocity dispersion measured  is also plotted.
The dashed lines represent the uncertainty, as measured by the
standard deviation if the values, on this mean.
The dotted line shows the true velocity dispersion of the model
cluster. 
See the online journal for a colour image. 
}
\end{figure}

The results are shown in Fig. \ref{fig:snr}. The mean of the
five values measured for each signal-to-noise ratio are also plotted,
and an uncertainty on this mean equivalent to the standard deviation
is represented by the dashed lines.
The measured velocity dispersion increases with decreasing
signal-to-noise ratio of the cluster, in agreement with the results
found by \citet{martini04}. 
The plot also shows that, in this case, a cluster signal-to-noise of
$\gtrsim\,2.1$ is  within $20\,\%$ of the true broadening, 
and the lower limit on the signal-to-noise ratio  for which the
standard deviation of the results describes the true error is just
over $3$. 
The precise signal-to-noise ratio at which these minima occur will
depend of the cluster and  on the nature of the template mismatch,
which should be less stark than that illustrated here. 
This plot indicates that the velocity dispersion measured for
cluster~805 is robust, although, as already stated, the standard
deviation of the results from the $\chi^2$ minimisation may
underestimate the true uncertainty. The larger
uncertainty measured from the standard deviation of all five templates
with both methods should remain a reliable estimate of uncertainty.

\subsection{Implications for cluster~805}

\begin{table}
\begin{center}
\caption{
\label{tab:param}
Parameters for cluster~805 taken from the literature and determined
in this work.}
\begin{tabular}{|l|r@{$\,\pm\,$}l|l|} \hline
 Parameter   &\multicolumn{2}{|c|}{Value}    & Reference\\ \hline
Distance (Mpc)                 &4.5   &0.3    &\citet{thim03}  \\
$M_V$ (mag)                    &--12.2 &0.4    &\citet{larsen04_rich} \\
Age (Myr)                      &13    &6      &\citet{larsen04_rich} \\
$M_\mathrm{phot}\,(\times\,10^5\,M_\odot)^\star$
                               &1.9   &1.4    &\citet{larsen04_rich}\\ 
$r_{\mathrm{hl}}$ (pc)         &2.8   &0.4    &\citet{larsen04_rich} \\
$\sigma_\mathrm{los}\,(\mathrm{km\,s}^{-1})$  
                               &10.2  &1.1    & This work \\
$M_\mathrm{vir}\,(\times\,10^5\,M_\odot)$ 
                               &6.6   &1.7    & This work \\ 
$L_V/M_\mathrm{vir}\,(L_\odot/M_\odot)$
                               &10    &5      & This work \\ \hline
\multicolumn{4}{l}{$^\star$assuming \cite{kroupa02} IMF.} \\
\end{tabular}
\end{center}
\end{table}

 Adopting the cluster parameters determined for cluster~805
  contained in Table \ref{tab:param},   and using the velocity
  dispersion measured here, we find
$M_\mathrm{vir}\,=\,(6.6\,\pm\,1.7)\,\times\,10^5\,\,M_\odot$
and a $V$-band luminosity-to-dynamical mass ratio
$L_V/M_\mathrm{vir}\,=\,(10\,\pm\,5)~L_\odot/M_\odot$.
These results agree, within the uncertainties, with the values
determined by \citet{larsen04_rich} of 
$M_\mathrm{vir}\,=\,(4.2\,\pm\,0.67)\,\times\,10^5\,\,M_\odot$  and 
$L_V/M_\mathrm{vir}\,=\,(15\,\pm\,6)~L_\odot/M_\odot$. Both of these
dynamical mass measurements are larger than the photometric
mass predicted for a \citet{kroupa02} IMF of
$M_\mathrm{phot}\,=\,(1.9\,\pm\,1.4)\,\times\,10^5\,\,M_\odot$ for the
cluster age of $13\,\pm\,6$~Myr
\citep{larsen04_rich}, indicating that cluster~805 may not be
virialised. However,  the photometric and dynamical masses agree at
just over the $2\sigma$ level.

Our value of light-to-dynamical-mass ratio suggests that the eSFE of
cluster~805 lies between $\sim 30\,\%$ and $\sim 50\,\%$, favouring
the value of $\sim 40\,\%$. \citet{goodwin06_bast} estimate that  a
cluster with an  eSFE~$\sim 30\%$ will lose $\sim 80\%$ of its mass
within $\sim40$~Myr  and disperse.  However, a cluster with
an  eSFE~$\gtrsim\,40\%$ will virialise, although it will lose up to
$\sim 60\%$  of it mass, and could 
  survive for a Hubble time in   the absence of external disruptive
  forces.


\section{Discussion: using  high-resolution, fibre-fed spectroscopy
  to determine velocity dispersions}
\label{sec:dis}

Using  high-resolution, fibre-fed spectroscopy
  to determine velocity dispersions  would have a major impact on
  producing a large sample of dynamical cluster mass determinations,
  by making more efficient use of telescope time. By having a large sample of
  dynamical cluster masses, it should become possible to assess how
  likely YMCs are to survive for a large fraction of a Hubble time and
  evolve into globular cluster-type objects.

The primary concern over this method is the question of how
successfully the galaxy background could be subtracted from
fibre spectroscopy of faint clusters. 
Of course, sky subtraction  is not a problem with bright stars, despite a small
amount of flux `bleeding' from the bright star fibres into the sky fibres.
For the fainter objects, galaxy background subtraction is more
challenging. Simply scaling and
subtracting the galaxy background fibre with the most similar
intensity background to the cluster being considered introduces  
more noise into the spectrum. Taking an average of two or
three scaled galaxy background fibres minimises this, although 
still introduces noise. Subtracting an average of two galaxy
background fibres decreases the signal-to-noise ratio of the
cluster\,+\,background spectrum by a factor of  $\sim 2.5$ for
cluster~805, the brightest cluster, and a  factor of $\sim 4$ for
both cluster~347, the faintest cluster, and cluster~645, which has the
lowest contrast with the galaxy background.
 It seems that scaling the galaxy background in this way is
a reasonable approximation to the local galaxy background.
All three galaxy background fibres are very similar, considering the
level of noise, in spite of 
spatial variations in galaxy background, due, for example, to
differing stellar populations and extinction.
Nevertheless, some sky lines are imperfectly subtracted in the
cluster~805 spectrum, although these are  
within the level of the noise.
 A reduced $\chi^2$ between the low- and mid-background spectra and
the scaled, high-background spectrum 
yields values of $\chi^2\,=\,0.9$ and $\chi^2\,=\,0.7$, respectively. 
While adopting the noise in the spectra as the uncertainty in the fit
seems to overestimate the true uncertainty, the relative similarity in the
galaxy background spectra, given the large degree of noise, is
demonstrated. 
An alternative method would be to fit the continuum of the  scaled, 
  extracted galaxy  background fibre and subtract this fit from the
  cluster spectrum, manually removing the sky lines  from the cluster
  spectrum.
 However, the extra complications involved in this process were
 not warranted for our data, since an appropriate galaxy background
 was achieved from averaged galaxy background fibres.

Also, the signal-to-noise ratio was much higher in the UVES spectrum of
  \citet{larsen04_rich}, for which a shorter integration time of
  7500\,s was used. 
This would appear to indicate that long-slit echelle spectroscopy
  is a more efficient use of telescope time, especially since two
  clusters could potentially be observed in the same slit. 
  However, finding  cases where two clusters can be observed
  simultaneously can be difficult, due to the relatively short slit
  length of many   echelle sepctrographs (e.g. 11\,arcsec for UVES).
The signal-to-noise ratio in the galaxy-subtracted
UVES+FLAMES cluster spectra was substantially lower than anticipated
from the ESO exposure time calculator. This predicted a
signal-to-noise ratio in excess of 30 at $\sim\,8000\,\textrm{\AA}$,
neglecting the high galaxy background. 
Even considering this additional noise does not account for the much
lower observed cluster signal-to-noise ratios.
The most likely explanation for the greatly reduced signal-to-noise
ratio is is that the fibres were positioned off-centre. However, this 
needs to be investigated further.

 As discussed in Section \ref{sec:comparison}, the 
signal-to-noise ratio of cluster~805 is a good representation of the
lowest signal-to-noise ratio required to measure robust
velocity dispersions. 
Therefore, making a conservative assumption that $50\%$ of the flux of
cluster~805 was missed in our spectrum indicates that results could be
obtained for a cluster half as bright as cluster~805 (i.e.,
$m_I\,\approx\,16.7$~mag). 
Approximately 18 of the 159 YMCs in M83, and 109 of all 1358 YMCs,
contained in the catalogue of  \cite{larsen04_cat} fulfil this criterion.

Another problem with this method, specific to FLAMES+UVES, is the
limited choice of spectral region, as 
dictated by the need to use  a standard UVES setup. This meant that
one of the strong lines of the Ca~{\sc ii} triplet was lost. Furthermore,
much of the data in the near-infrared cannot be used to determine
velocity dispersions, due to
 telluric features and the lower CCD response in this spectral region. 
  Clusters also need to be selected carefully, with
$M_\mathrm{vir}\,\gtrsim\,10^5\,M_\odot$ in order to have 
high enough velocity dispersions to be measured. 
 For a 10-Myr-old cluster, this corresponds to
  $M_V  = -11.5\,\mathrm{mag}$   ({\sc starburst99};
  \citealt{leitherer99}),  or $m_V \sim 16.8\,\mathrm{mag}$ for a
  distance of 4~Mpc 
  and $E(B-V)\,\sim\,0.1$\,mag.

While the drawbacks of this method must still be borne in mind, we
have shown here that reliable velocity dispersions can be measured
from fibre-fed spectroscopy. Therefore, this
method can potentially be used to obtain a large sample of dynamical
masses, which can be used to address the issue of the nature of YMCs
as proto-globular clusters.


\section{Summary}
\label{sec:sum}

We determined a velocity dispersion of cluster~805 in M83 of
$\sigma_\mathrm{los}\,=\,10.2\,\pm\,1.1\,\mathrm{km\,s}^{-1}$ from
VLT/FLAMES+UVES  
spectroscopy. 
This agrees well with the results obtained here for the standard
echelle VLT/UVES spectrum of 
cluster~805 from  \citet{larsen04_rich} of
$\sigma_\mathrm{los}\,=\,10.6\,\pm\,1.4\,\mathrm{km\,s}^{-1}$.
This shows that high-resolution,
fibre-fed spectroscopy is a practicable method with which to
measure the velocity dispersions of several clusters
simultaneously. 
This  allows a more efficient use of
telescope observing time, providing that sufficiently high
signal-to-noise ratios  are achieved. 
Therefore, high-number statistics of YMC dynamical masses, vital
for assessing the role of YMCs as proto-globular clusters can be more 
easily obtained. 

We find no systematic differences in the velocity dispersions
measured depending on whether red giant or red supergiant templates
were considered. Nor do we find any differences in the velocity
dispersions determined from the $\chi^2$ technique when the Ca~{\sc
  ii} triplet lines are included in the fit. While the results produced
from both a $\chi^2$ minimisation and from cross-correlation are
consistent, there are indications that adopting the
standard deviation of the results determined from only one technique
may underestimate the true uncertainties.

Adopting our value of velocity dispersion and all other cluster
parameters from \citet{larsen04_rich}, we measure a virial mass of
$M_\mathrm{vir}\,=\,(6.6\,\pm\,1.7)\,\times\,10^5\,M_\odot$ for
cluster~805,  somewhat larger
than its photometric mass of 
$M_\mathrm{phot}\,=\,(1.9\,\pm\,1.4)\,\times\,10^5\,M_\odot$, for a
\citet{kroupa02} IMF and an age of $13\,\pm\,6$~Myr. Combining
this  with the eSFE predictions of \citet{goodwin06_bast}, we conclude
that the cluster will probably virialise and could survive for a Hubble time in
the absence of external disruptive forces.


\section*{Acknowledgements}

We would  like to thank Arianne Lan\c{c}on for providing model
red supergiant spectra to help select the spectral regions over which
to determine the velocity dispersion. 
We also appreciate suggestions made by the anonymous referee which
have helped to improve this manuscript. 
The Image Reduction and Analysis Facility {\sc iraf} is distributed by
the National Optical Astronomy Observatories, which is operated by the
Association of Universities for Research in Astronomy, Inc., under
cooperative agreement with the U.S. National Science Foundation. 
SLM acknowledges financial support from STFC. 

\bibliographystyle{aa}
\bibliography{abbrev,refs}

\end{document}